\documentclass[aps,prl,floats,floatfix,twocolumn]{revtex4}
\usepackage{color}
\usepackage{bm}
\usepackage{latexsym}
\usepackage{amsmath}
\usepackage[pdftex]{graphicx}

\begin{document}

\newcommand{\cc}{{\bf\Large C }}
\newcommand{\half}{\mbox{\small $\frac{1}{2}$}}
\newcommand{\sinc}{\mbox{sinc}}
\newcommand{\trc}{\mbox{trace}}
\newcommand{\intt}{\int\!\!\!\!\int }
\newcommand{\ointt}{\int\!\!\!\!\int\!\!\!\!\!\circ\ }
\newcommand{\ar}{\mathsf r}
\newcommand{\bmsf}[1]{\bm{\mathsf{#1}}}
\newcommand{\dd}[1]{\:\mbox{d}#1}

\newcommand{\ham}{\mathcal{H}}
\newcommand{\PT}{\mathcal{PT}}
\newcommand{\const}{\mbox{const.}}
\newcommand{\EPS} {\mbox{\LARGE $\epsilon$}}

\newcommand{\hide}[1]{}
\newcommand{\tbox}[1]{\mbox{\tiny #1}}
\newcommand{\mbf}[1]{{\mathbf #1}}

\newcommand{\im}[1]{\mbox{Im}\left\{#1\right\}}
\newcommand{\re}[1]{\mbox{Re}\left\{#1\right\}}
\newcommand{\abs}[1]{\left|#1\right|}
\newcommand{\bra}[1]{\left\langle #1\right|}
\newcommand{\ket}[1]{\left|#1\right\rangle }
\newcommand{\eexp}{\mbox{e}^}

\newcommand{\eq}[1]{Eq.~(\ref{eq:#1})}
\newcommand{\fig}[1]{Fig.~\ref{fig:#1}}
\newcommand{\eos}{\,.}
\definecolor{red}{rgb}{1,0.0,0.0}


\title{$\PT$-symmetric Wave Chaos}
\author{Carl T. West$^{1,2}$, Tsampikos Kottos$^{1}$, and Toma\v z Prosen$^{3}$}

\affiliation{
$^1$Department of Physics, Wesleyan University, Middletown, Connecticut 06459, USA\\
$^2$Max-Planck-Institute for Dynamics and Self-Organization, 37073G\"ottingen, Germany\\
$^3$Physics Department, Faculty of Mathematics and Physics, University of Ljubljana, Ljubljana, Slovenia
}

\date{\today}

\begin{abstract}
We study a new class of chaotic systems with dynamical localization, where gain/loss mechanisms break 
the Hermiticity, while allowing for parity-time ${\cal PT}$ symmetry. For a value $\gamma_{\cal PT}$ 
of the gain/loss parameter the spectrum undergoes a spontaneous phase transition from real (exact phase) 
to complex values (broken phase). We develop a one parameter scaling theory for $\gamma_{\cal PT}$, 
and show that chaos assists the exact ${\cal PT}$-phase.  Our results have applications to the 
design of optical elements with ${\cal PT}$-symmetry.
\end{abstract}

\maketitle


$\mbf{{}Introduction. -}$ 
Systems exhibiting parity-time (${\cal PT}$) symmetry have been the subject of rather intense research 
activity during the last few years. This interest was motivated by various areas of physics, ranging 
from quantum field theories and mathematical physics \cite{BB98,BBM99,BBJ02,B07} to solid state physics
\cite{H92,BFKS09} and classical optics \cite{GMCM07,RDM05,MGCM08,MMGC08,L09,GSDMVASC09,Secret}. A surprising 
result that was pointed out in some of these investigations was the possibility that ${\cal PT}$ symmetric 
hamiltonians ${\cal H}$ can have real spectrum, despite the fact that they can, in general, be
non-Hermitian \cite{B07}. The departure from Hermiticity, is due to the presence of various gain/loss mechanisms 
which occur in a balanced manner,  
so that the net loss or gain of ``particles" is zero. Furthermore, as some gain/loss parameter $\gamma$ that 
controls the degree of non-hermiticity of ${\cal H}$ changes, a spontaneous ${\cal PT}$ symmetry breaking 
occurs. At this point, $\gamma=\gamma_{\cal PT}$, the eigenfunctions of ${\cal H}$ cease to be eigenfunctions 
of the ${\cal PT}$-operator, despite the fact that ${\cal H}$ and the $\mathcal{PT}$-operator commute \cite{B07}. 
This happens because the ${\cal PT}$-operator is anti-linear, and thus the eigenstates of ${\cal H}$ may or 
may not be eigenstates of ${\cal PT}$. As a consequence, in the {\it broken $\cal{PT}$ -symmetry phase} the 
spectrum becomes partially or completely complex. The other limiting case where both ${\cal H}$ and ${\cal 
PT}$ share the same set of eigenvectors, corresponds to the so-called {\it exact $\mathcal{PT}$-symmetric 
phase} in which the spectrum is real.

A promising realization of ${\cal PT}$ symmetric systems appears in the frame of optics, where a medium 
with alternating regions of gain and loss can be synthesized, such that the (complex) refraction index 
satisfies the condition $n^*(-x)=n(x)$ \cite{BFKS09,GMCM07,MGCM08,MMGC08,L09,RDM05}. This kind of synthetic 
${\cal PT}$ materials exhibits unique characteristics such as ``double refraction" and non-reciprocal 
diffraction patterns, which may allow their use as a new generation of unidirectional optical couplers 
or left-right sensors of propagating light \cite{MGCM08}. Recently, the interest in ${\cal PT}$ systems 
bursted further 
due to their experimental realization \cite{GSDMVASC09,Secret}. In this respect, one of the emerging 
questions is how one can enhance the parameter regime for which exact ${\cal PT}$-phase is present, while 
at the same time provide a general theoretical formalism for the behavior of $\gamma_{\cal PT}$, in terms 
of system parameters like imperfections, system size, complexity of the underlying classical (ray) dynamics 
etc.
 
In this Letter, we investigate the behavior of the exact ${\cal PT}$-phase in a new setting of systems, 
namely a class of Hamiltonians whose classical (ray) dynamics is chaotic while its quantum/wave analogue 
can show dynamical localization \cite{I90,FGP82}, a dual phenomenon to Anderson localization appearing 
in disordered media \cite{A58}. As a result of this duality, our study (although performed in the framework
of wave chaos systems) is directly relevant to disordered quasi-one dimensional systems like disordered 
arrays of optical fibers \cite{PPKSBNTL04,LSCASS08}. We have developed a {\it one parameter scaling theory} 
for $\gamma_{\cal PT}$ and show that it is the only relevant parameter that controls the variation of 
$\gamma_{\cal PT}$ with $N$, the system size of a sample. Specifically,
\begin{equation}
  \label{eq:Scaling}
  \frac{\partial {\tilde \gamma}_{\PT}}{\partial \log N} = \beta({\tilde \gamma}_{\PT});\quad {\rm where} \quad
{\tilde \gamma}_{\PT}\equiv N\gamma_{\cal PT}
\end{equation}
where $\beta$ is a {\it universal} function of ${\tilde \gamma}_{\PT}$ alone.

Furthermore, we have investigated the distribution of the $\PT$-parameter ${\cal P}(\gamma_{\PT})$ in the 
localized and delocalized/chaotic regimes and found that it reflects the properties of the respective 
system. Specifically, we have found that ${\cal P}(\gamma_{\PT})$ is $\log$-normal in the former case,
while in the latter it follows a Wigner distribution, reflecting the level repulsion characterizing 
systems with chaotic/diffusive dynamics. Our results have direct applications not only to coupled optical 
${\cal PT}$-elements but also to cold atoms moving in a complex ${\cal PT}$-potential \cite{PTatoms}.

$\mbf{{}Model. -}$
The prototype model of quantum chaos is the celebrated Kicked Rotor (KR) which exhibits the phenomenon 
of dynamical localization (DL) \cite{I90,FGP82}. Specifically, it has been shown that the quantum 
suppression of classical diffusion taking place in momentum space is a result of wave interference 
phenomena similar in nature to the ones responsible for Anderson localization in random media. We study
a variation of the Kicked Rotor (the $\PT$KR) \cite{KR1,I90,CGGI94,KR_DL}, defined by the time-dependent
 Hamiltonian \cite{note1}
\begin{equation}
  \label{eq:PTKR_Ham}
  \ham = \frac{p^2}{2} + K_0 V(q) \sum_n\delta(t-nT);\,\,V(q)=\cos(q)+i\gamma q
\end{equation}
where $(q,p)$ are a pair of canonical variables and $q \in [-\pi,\pi)$ \cite{WhyLinear}. The kicks have 
strength $K_0$ and period $T$, which we set to unity without loss of generality. In order to avoid any 
integrable regions in the classical phase space (at $\gamma=0$) we take $K_0\ge 5$. We mark that in the 
framework of geometric optics Eq.~(\ref{eq:PTKR_Ham}) describes the propagation of light ray along a chain 
of optical elements equally placed along the axis of propagation $t$, in distance $T$ from one-another 
\cite{K87,PF89}. Furthermore we assume that the elements are purely refractive and ideally thin 
with a variation only in one transverse direction $q$. The phase space variable $p=n_0 {\rm d}q/{\rm d}t$ 
is proportional to the slope of the ray, while $n_0$ is the free-space refractive index. 

The wave (quantum) dynamics of this system is described by the one-period evolution operator
\begin{equation}
  \label{eq:PTKR_Opr}
  \mathcal{U}=\exp(-i\hat{p}^2/4\hbar)\exp(-ikV(\hat{q}))\exp(-i\hat{p}^2/4\hbar)
\end{equation}
where $k=K_0/\hbar$, ${\hat p}=\hbar {\hat l}= -i\hbar {\rm d}/{\rm d}q$ and $-N/2\leq l\leq N/2$. For 
$\gamma=0$, it was found that the eigenfunctions $\psi_l\equiv \langle l|\psi\rangle$ of ${\mathcal U}$, 
are exponentially localized in momentum space with a localization length $\xi \equiv \lim_{N\rightarrow
\infty}1/\sum_l^N |\psi_l|^4 \approx k^2/2$ \cite{I90,FGP82}. They are solutions of the eigenvalue problem 
\begin{equation}
\label{eigen}
\mathcal{U}\ket{\psi} = \lambda \ket{\psi};\quad \lambda=\exp(-i\epsilon)
\end{equation}
where the eigenvalues $\lambda$ are unimodular at $\gamma=0$, and the phases $\epsilon$ are referred to as 
quasi-energies. In the case of $\hbar=2\pi M/N$ with $M$,$N$ integers, Eq.~(\ref{eq:PTKR_Opr}) defines a dynamical 
system on a torus. The localization properties of the eigenstates are determined by the scaling parameter 
$\Lambda= N/\xi$: if $\Lambda\gg 1$ the eigenstates are exponentially localized while if $\Lambda \ll 1$ 
they are ergodically spread over the momentum space.

\begin{figure}[h]
\includegraphics[width=1.0\columnwidth,keepaspectratio,clip]{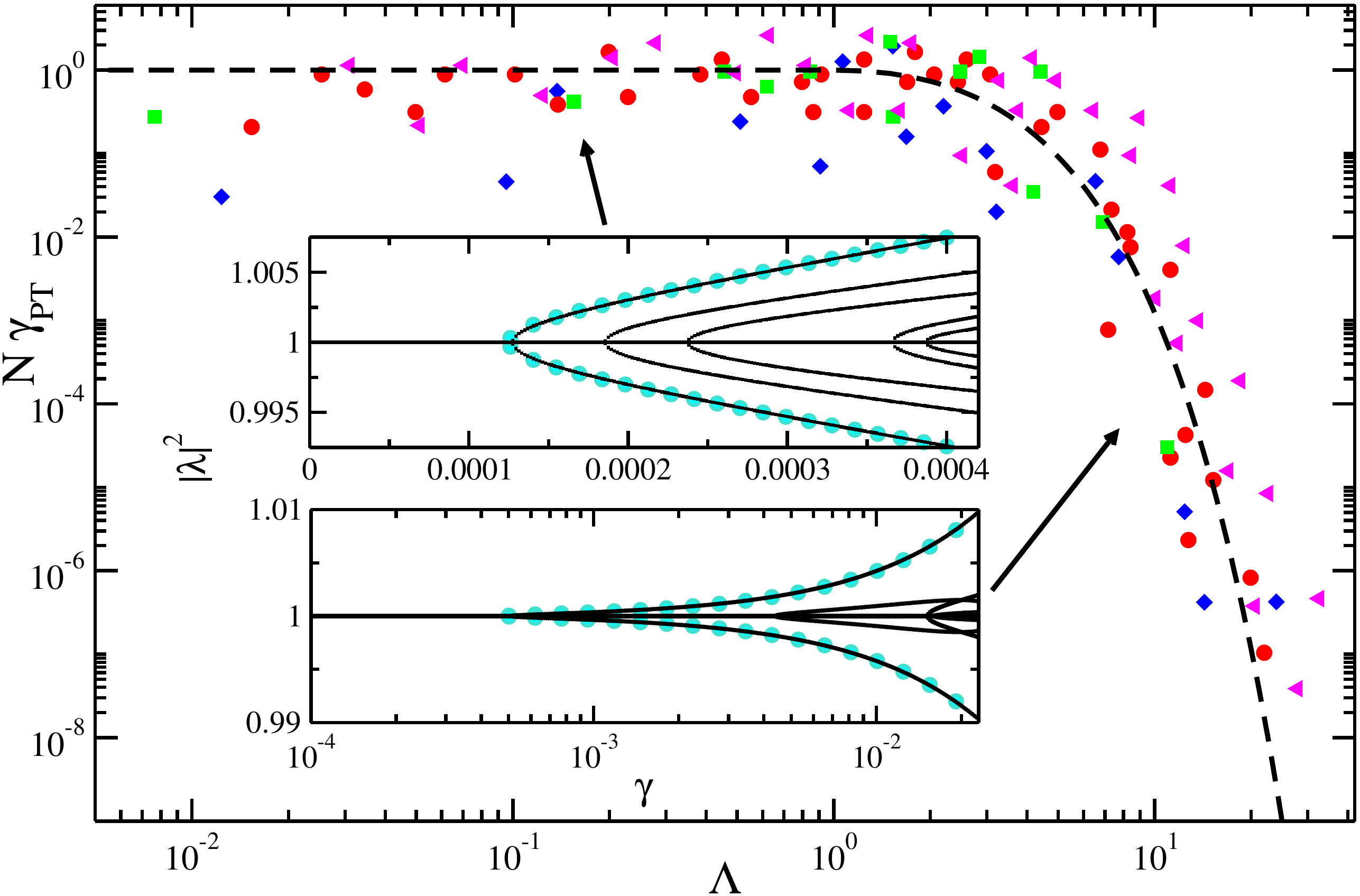}
 \caption{
   \label{fig:Scaling_Splitting} 
   (color online) Scaling behavior of $\gamma_{\PT}$ for the $\PT$KR defined by
   \eq{PTKR_Opr} for various $M$, and $K_0>5$ with $N=63$ (pink triangles),
   127 (red circles), 225 (green squares), and 511 (blue diamonds). All data
   shows nice collapse to the theoretical prediction in \eq{Scaling2} (dashed
   black line). Inset: Parametric evolution of eigenvalue magnitudes for
   Chaos (D.L.) in the upper (lower) figure with $N=127$, and $\Lambda=0.05$
   (10). The first branching pair is responsible for the transition to
   broken $\PT$ phase, and in both cases follows the predicted functional
   form (see text) shown with blue circles.}
\end{figure}

${\cal PT}-\mbf{{}breaking} \, \mbf{{}scenario}-$ 
In the exact $\PT$ phase (i.e. $\gamma\leq \gamma_{\cal PT}$) all eigenvalues are restricted to the unit 
circle, resulting in a real quasi-energy spectrum (see insets of Fig.\ref{fig:Scaling_Splitting}). We find 
that the mechanism for transition to the broken $\PT$ phase is a level crossing between the pair of eigenvalues
which are closest on the unit circle for $\gamma=0$. However, rather than splitting into the complex plane 
symmetrically around the real line as in the case of Hamiltonian systems \cite{BFKS09}, these pairs split 
in a logarithmically symmetric manner away from the unit circle (one to the interior, the other to the exterior).  
Considering the first breaking pair of levels as an isolated two level system \cite{2level} we find that the first branching 
is described by $|\lambda_{\pm}|^2 \propto \exp(\pm2\sqrt{ \gamma_{\PT}^2 -\gamma^2})$. This square root 
singularity near the bifurcation point is quite universal of an exceptional point and applies both for $\Lambda
\gg 1$ and $\Lambda\ll 1$. This is further confirmed numerically in \fig{Scaling_Splitting} where we present 
the spontaneous ${\cal PT}$-symmetry breaking scenario for two representative cases associated with
localized and delocalized/chaotic parameter values.

$\mbf{{}Scaling}\,  \mbf{{}theory} \, \mbf{{}for}\, \gamma_{\cal PT} -$ We consider first the limiting case 
$\Lambda\gg 1$ where dynamical localization is dominant. To clarify the picture we start from the Hermitian 
limit $\gamma=0$. Imagine for the moment that the kicking strength $k$ is zero. Then all states are 
$\delta-$like functions localized at various momenta $-N/2\leq l\leq N/2$. There is an exact degeneracy of 
multiplicity two between the states localized at $\pm l_0$, i.e. symmetrically around $l=0$. For $k\neq 0$ 
this degeneracy is lifted. The eigenstates whose centers of localization are a distance $d=2l_0\gg \xi\sim k^2$ 
apart, form a quasi-degenerate pair of symmetric /antisymmetric states \cite{CGGI94}. Each has two peaks, 
near the momenta $\pm l_0$, and decays exponentially $\psi(l)\sim (1/\sqrt{\xi}) \exp(-|\pm l_0-l|/\xi)$ away 
from them (double hump states). Thus, the eigenstates in a $\mathcal{P}$ -symmetric KR are organized into pairs 
(doublets) ordered by quasi-energy difference, $\delta_1< \delta_2<\cdots$. The splitting between quasi-degenerate 
levels is $\delta_{l_0}\sim (1/\xi) \exp(-2 l_0/\xi)$ while the energy separation between consecutive doublets 
is much larger, of the order of the mean level spacing of the system, $\Delta \sim 1/N$. Specifically, the 
smallest energy splitting $\Delta_{\rm min}=\delta_1\sim (1/\xi)\exp(-N/\xi)$ corresponds to states that 
where originally located at the extreme points of the momentum ``lattice", i.e. $l_0=\pm N/4$. 

As $\gamma$ is switched on the eigenstates of each pair will initially preserve their ${\cal PT}$-symmetric 
structure \cite{BFKS09,LongPaper}. At $\gamma =\gamma_{\cal PT} \simeq \Delta_{\rm min}= \delta_{1}$, the two 
levels associated with $\delta_{1}$ will cross, breaking the $\mathcal{PT}$-symmetry (see Fig. 1 inset). As 
$\gamma> \gamma_{\cal PT}$ these modes cease to be eigenstates of the ${\cal PT}$-operator. Instead, the weight
of each is gradually shifted towards one of the localization centers \cite{BFKS09,LongPaper}. For larger $\gamma$
the next doublet (with splitting $\delta_2$) will come into play, creating a second pair of complex eigenvalues 
for $\gamma \simeq \delta_2$ (see Fig. 1 inset), etc.

Let us now consider the opposite limit of $\Lambda \ll 1$, where the eigenstates are ergodically spread all over 
the system. In this case, the picture of doublets with exponentially small energy splittings is not valid and 
$\gamma_{\cal PT}$ becomes of the order of the minimal level spacing, $\Delta_{\rm min}$, in the corresponding
Hermitian problem. This statement follows from perturbation theory with respect to $\gamma$. The unperturbed 
(i.e. $\gamma=0$) energy levels are real, and are separated by intervals of order $1/N$, so that $\Delta_{\rm min}
\simeq 1/N$. Finite $\gamma$ leads to level shifts proportional to $\gamma^2$ (the first order correction vanishes 
due to ${\cal PT}$-symmetry) and for $\gamma=\gamma_{\cal PT}\simeq\Delta_{\rm min}$ the perturbation theory 
breaks down, signaling level crossing and the appearance of the first pair of complex eigenvalues. Thus, the 
energy scale for the ${\cal PT}$-threshold in the $N/\xi \ll 1$ limit ($\gamma_{\cal PT}\simeq 1/N$) widely 
differs from that for $N/\xi \gg 1$ ($\gamma_{\cal PT} \simeq (1/\xi)e^{-N/\xi}$). 

Combining both cases, we conclude that in the two limits of weak and strong localization we have that
\begin{equation}
\label{sfunction}
{\tilde \gamma}_{\cal PT} = f(\Lambda)=\left\{
  \begin{array}{lcr}
    \mathcal{O}(1)                & {\rm for} & \Lambda \le 1 \\
    \Lambda\exp(-\Lambda) & {\rm for} & \Lambda >1
  \end{array}\right.
\end{equation}
Our numerical results for the $\PT$KT model, are reported in \fig{Scaling_Splitting} and are in excellent 
agreement with the above theoretical predictions. Moreover, they clearly show that the scaling function
$f(x)$ is regular and interpolates smoothly between the two limiting cases. This allow us to conclude that
\begin{equation}
  \label{eq:Scaling2}
  {\tilde \gamma}_{\PT}\equiv N\gamma_{\PT}=f(\Lambda\equiv {\frac{N}{\xi}}); 
\end{equation}
which can be rewritten in the form of Eq.~(\ref{eq:Scaling}). This is the main result of the present Letter,
as it allow us to postulate the existence of a $\beta$-function for the ${\tilde \gamma}_{\cal PT}$ of {\it
generic} chaotic (or quasi-$1D$ disordered) systems. In the remainder of the Letter, we will focus on the 
statistical properties of $\gamma_{\cal PT}$ as a function of the parameter $\Lambda$.

$\mbf{{}Distribution}\, \mbf{{}of}\, \gamma_{\PT}\mbf{. -}$ The above discussion pertain only to the behavior 
of a `typical' system. A full theory however must be formulated in statistical terms and deal with probability  
distributions ${\cal P}(\gamma_{\cal PT})$. To this end we exploit the equivalence between $\gamma_{\PT}$ and 
$\Delta_{\text{min}}$ which is confirmed numerically in the inset of Fig. 2a for $\Lambda$ values spanning the
whole interval from the localized to extended regime. Thus, we instead analyze the distribution 
${\cal P}(\Delta_{\text{min}})$. For better statistics, an ensemble of ${\cal P}$-symmetric KR systems has 
been created by randomizing the phases of the kinetic part of the evolution operator given by Eq.~(\ref{eq:PTKR_Opr}). 
In all cases, the numerical distribution ${\cal P}(\Delta_{\text{min}})$ involved more that $10^4$ data points for statistical 
processing. 

\begin{figure}[h]
\includegraphics[width=1.0\columnwidth,keepaspectratio,clip]{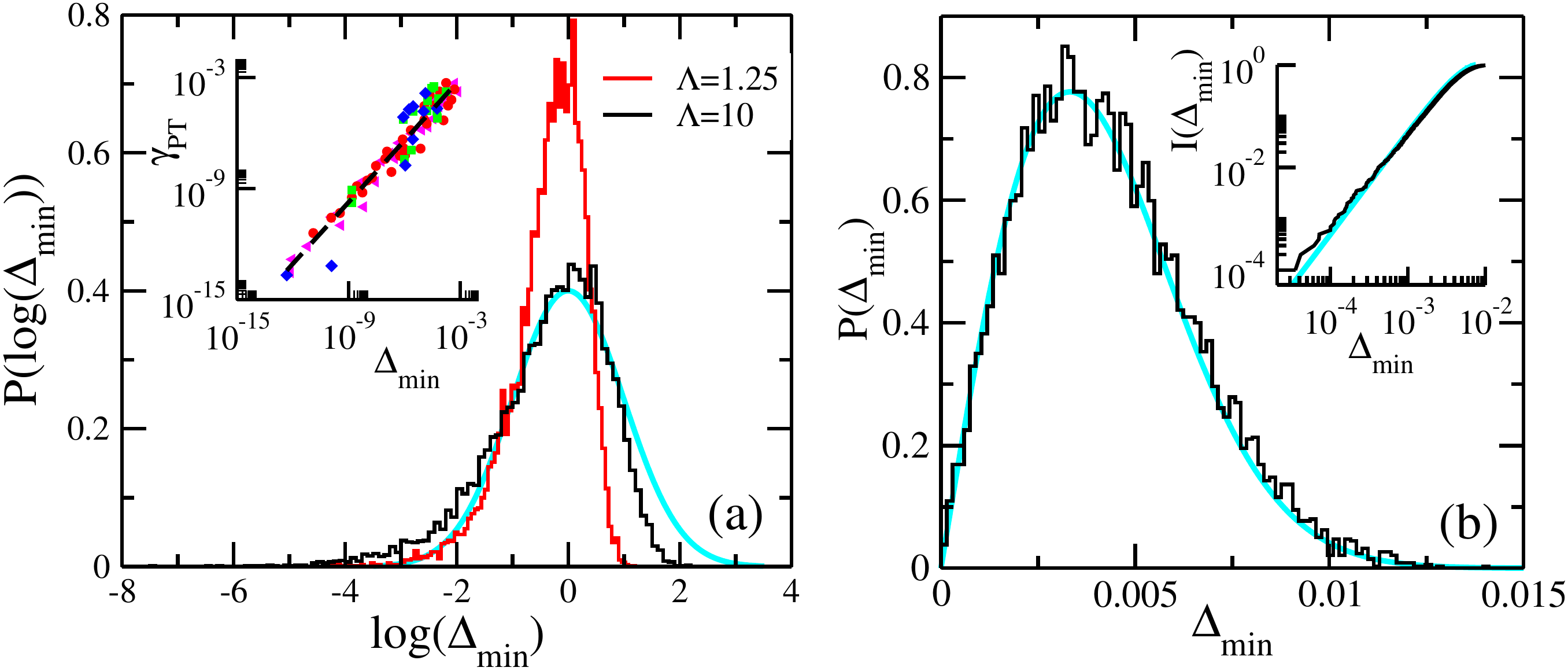}
 \caption{
   \label{fig:Distributions}
   (color online) (a): Distributions of $\Delta_{\rm min}$ for localized eigenfunctions displaying a convergence 
   to log-normal behavior. Centers are shifted for ease of comparison. Inset: Linear relation between 
   $\Delta_{\text{min}}$ and $\gamma_{PT}$ over roughly $12$ orders of   magnitude. Parameters and symbols are the 
   same as in \fig{Scaling_Splitting}. (b): Distribution of minimum level spacings in the chaotic regime in which
   Wignerian behavior (blue line) is observed. Inset: The integrated distribution $I(\Delta_{\rm min})= 
   \int_0^{\Delta_{\rm min}} {\cal P}(x) dx$ in a log-log plot. The best fit (blue line) has power two. In both 
   cases, $N=127$. 
 }
\end{figure}

We start our analysis from the localized regime $\Lambda\gg 1$. There are several sources of fluctuations in 
$\delta_{1}$: fluctuations in the position and energy of the relevant localized states, as well as what can be 
termed ``fluctuations in the wave functions''. By this we mean that a localized wave function exhibits strong, 
$\log$-normal fluctuations in its ``tails'', i.e. sufficiently far from its localization center \cite{TailFluctuations}.
This latter source of fluctuations appears to be the dominant one and it immediately yields a log-normal distribution 
for $\delta_{1}$ (see \fig{Distributions}), since $\delta_1$ is proportional to the overlap integral between 
a pair of widely separated and strongly localized states \cite{TailFluctuations,BFKS09}. We confirm numerically 
that for increased $\Lambda$ we do in fact approach such a distribution (see \fig{Distributions}). 

In contrast, in the delocalized regime $\Lambda\ll 1$, the structure of the eigenfunctions is random \cite{I90,CGGI94}
and they are ergodically spread over the momentum space. Thus there is no quasi-degeneracy for such states; 
instead, there is a level repulsion due to strong eigenstate overlap. This results in a Wignerian distribution 
for the minimum energy difference, i.e. ${\cal P}(\Delta_{\rm min})=(\pi/2)\Delta_{\rm min} \exp(-\pi \Delta_{
\rm min}^2/4)$, which is dramatically different from the one found in the localized regime (see \fig{Distributions}a). 
Since ${\cal P}(\Delta_{\rm min}) \sim {\cal P}(\gamma_{\cal PT})$ we conclude that in the case of wave chaos 
${\cal P}(\gamma_{\cal PT}\rightarrow 0 ) \rightarrow 0$ i.e. there always exists a $\gamma$-interval for which 
we will have an exact ${\cal PT}$ -phase. This observation can be used as a new criterion of wave chaos. 
Our numerical results for $\Lambda=0.01$ are shown in \fig{Distributions}b, and are in excellent agreement with 
the above theoretical considerations. 

{\it Conclusions -}
In conclusion, we have studied a new class of quantum chaotic systems with dynamical localization that 
also possess a ${\cal PT}$-symmetry. These systems are described by a non-Hermitian hamiltonian due to 
the existence of well-balanced gain/loss mechanisms and show a spontaneous ${\cal PT}$-symmetry breaking, 
i.e. a transition from a real to a complex spectrum, for some value $\gamma_{\cal PT}$ of the gain/loss 
parameter $\gamma$. We have developed a one parameter scaling theory for the rescaled critical gain/loss 
parameter ${\tilde \gamma}_{\cal PT}=\gamma_{\cal PT}N$, and conclude that there is a {\it universal} 
$\beta$-function that depends only on ${\tilde \gamma}_{\cal PT}$ itself, which controls the variation
of ${\tilde \gamma}_{\cal PT}$ with the system size. Furthermore, we have analyzed the distribution
${\cal P}(\gamma_{\cal PT})$ in the localized/delocalized regimes, and show that it drastically 
differs in these two limits. In the former case it is log-normal while in the latter it
follows a Wigner statistics reflecting the chaoticity of the underlying classical dynamics.
Our study opens the way to quantify the spontaneous  breaking of the $\PT$-symmetry in terms
of {\em universal} $\beta$-function. 

The results presented here are based on a simple connection between $\gamma_{\PT}$ and the (minimal) 
level spacing $\Delta_{\rm min}$ (see inset of Fig. \ref{fig:Distributions}a) which is inspired by an 
isolated two level matrix model (see section on the ${\cal PT}$-breaking scenario). Although 
this is the most generic type of scenario, describing a large number of physical realizations associated 
with classically chaotic (or quasi-$1D$ disordered) systems, there are many other interesting cases
which needs to be explored. For example, in the $\PT$-symmetric KR model presented above, we used a 
discontinuous (at the boundaries) imaginary potential ${\rm Im}V(q)$ which gives a lower bound on 
the behaviour of $\gamma_{\PT}$ (worst case scenario). However, preliminary analysis 
\cite{LongPaper} shows that if ${\rm Im}V(q)$ is a continuous and analytic function the ${\cal PT}$-symmetry breaking
scenario can be different. In such a case the matrix elements of ${\rm Im} V(q)$ between ${\cal P}$-
doublets are exponentially small in the regime of dynamical localization. Thus, according to the lowest 
order perturbation theory, the energy levels of these two states remain nearly parallel (as a function 
of $\gamma$), and hence typically $\gamma_{\rm PT} \gg \Delta_{\rm min}=\delta_1$. We also note that 
one can observe {\em spontaneous anti-breaking} of ${\cal PT}$-symmetry, i.e. for some $\gamma \gg 
\gamma_{\rm PT}$, a pair of non-unimodular eigenvalues recombines again into a pair of uni-modular 
eigenvalues, and sometimes, (but more rarely as $N$ is increasing) one may even find situations for 
which all levels simountaneously become uni-modular (global recovery of the exact ${\cal PT}$-phase)
\cite{LongPaper}. Understanding of such anti-breaking mechanism could be of significant interest for 
optics applications. Because of lack of space these results will be discuss elsewhere \cite{LongPaper}.
 
It will be interesting to extend this line of study to higher dimensions (possibly $3D$ disordered 
systems with a metal-to-insulator phase transition). We expect that our study will be of interest not 
only for the optics community but also for the atomic physics community where
complex optical potentials have been recently constructed \cite{PTatoms}.

We acknowledge usefull discussions with B. Shapiro and R. Fleischmann. This work was supported by the 
DFG FOR760, a grant from the US-Israel Binational Science Foundation (BSF), Jerusalem, Israel, and grant 
P1-0044 of Slovenian Research Agency.  CTW acknowledges the hospitality of FMF, University of Ljubljana. 


\begin{thebibliography}{99}
\bibitem{note1} We mark that the extension to non-autonomous $\PT$ systems is another
novelty of our calculations, which is of potential interest to the cold atoms community.
\bibitem{WhyLinear} If the imaginary component of the potential it taken to be
  continuous and analytic at $q=\pm \pi$, then coupling matrix elements of ${\rm Im} V(q)$ between 
  localized states in momentum space are exponentially small and one has a fundamentally different behavior as discussed later.
\bibitem{2level} We argue, in the spirit of degenerate perturbation theory, that the two level
approximation becomes accurate as soon as the level splitting between the obseved two levels is much smaller
than all the level spacing to other levels.
\end{thebibliography}
\end{document}